\documentclass[12pt]{iopart}
\usepackage{graphicx, dcolumn, iopams}
\usepackage{MjEpp, upgreek, tikz, dsfont}
\usepackage[utf8]{inputenc}
\usepackage[breaklinks=true, colorlinks=true, linkcolor=blue, urlcolor=blue, citecolor=blue]{hyperref}

\usetikzlibrary{positioning,shapes}

\def\DO         {\hat{\rho}}
\def\HWPi       {\hat{\Pi}_f}
\def\OpPi       {\hat{\Pi}}
\def\HWPar      {\alpha}
\def\SpinPi     {\hat{\Pi}_a}
\def\SpinPar    {\theta,\phi}

\begin{document}

\title{Visualization of correlations in hybrid quantum systems}

\author{R~P~Rundle$^{1,2}$, B~I~Davies$^1$, V~M~Dwyer$^{1,2}$, Todd Tilma$^{3,4,1}$ and M~J~Everitt$^1$}
\address{$^1$ Quantum Systems Engineering Research Group, Department of Physics, Loughborough University, Leicestershire LE11 3TU, United Kingdom}
\address{$^2$ The Wolfson School, Loughborough University}
\address{$^3$ Department of Physics, College of Science, Tokyo Institute of Technology, H-63, 2-12-1 \=Ookayama, Meguro-ku, Tokyo 152-8550, Japan}
\address{$^4$ Quantum Computing Unit, Institute of Innovative Research, Tokyo Institute of Technology, S1-16, 4259 Nagatsuta-cho, Midori-ku, Yokohama 226-8503, Japan}
\eads{\mailto{m.j.everitt@physics.org}}

\date{\today}

\begin{abstract}
    In this work we construct Wigner functions for hybrid continuous and discrete variable quantum systems. 
    We demonstrate new capabilities in the visualization of the interactions and correlations within hybrid quantum systems.
    Specifically, we show how to clearly distinguish signatures that arise due to quantum and classical correlations in an entangled Bell-cat state.
    We further show how correlations are manifested in different types of interaction, leading to a deeper understanding of how quantum information is shared between two subsystems.
    Understanding the nature of the correlations between systems is central to harnessing quantum effects for information processing; the methods presented here reveal the nature of these correlations, allowing a clear visualization of the quantum information present in hybrid quantum systems.
    The methods presented here could be viewed as a form of quantum state spectroscopy.
\end{abstract}

\maketitle

\section{Introduction}
Quantum correlations have become central to the design and manufacture of various quantum technologies~\cite{WheelerBook,NielsonChuang,WisemanMilburn,GerryKnight}.
Whether these quantum correlations are found between macroscopically distinct superpositions of states, also known as Schr\"odinger cat states, or in the entanglement between multiple systems.
Currently, such technologies can be broadly categorized as being based on either continuous-variable (CV) or discrete-variable (DV) quantum systems.

For CV systems, the primary focus has been on quantum optical systems; manipulating coherent states of light for various quantum information processing applications~\cite{Ralph2003,Gilchrist2004,RevModPhys.77.513,Neergaard-Nielsen2010}.
In such systems, the Wigner function~\cite{Wigner1984,Scully1997} is commonly used due to its ability to display an intuitive representation of a quantum state.
Furthermore, the Wigner function is particularly good at revealing coherences and correlations, such as squeezing and superposition~\cite{Breitenbach}.
For these reasons, it has become a fundamental tool in the `search' for Schr\"odinger’s cats~\cite{Brune1992}, readily identified by the iconic interference patterns arising from its quantum correlations.

By contrast the focus for DV systems has been on exploiting two-level quantum systems -- \textit{qubits} -- in order to generate a quantum analogue of the classical bit~\cite{NielsonChuang, Schumacher1995, Ladd2010}. 
Here, the Wigner function has received little attention as a means of visualization.
Unlike the case of CV systems, there are two common approaches for generating informationally complete DV Wigner functions, both of which have found application.
The approach developed in~\Ref{WOOTTERS19871, PhysRevA.70.062101} uses discrete degrees of freedom and has proven useful for quantum information purposes, particularly in the case of contextuality and Wigner function negativity~\cite{Howard2014,Delfosse2015,Raussendorf2017}.
The second approach (and the one used in this work) uses a DV Wigner function with continuous degrees of freedom, similar to the Bloch sphere~\cite{PhysRevA.49.4101, Varilly:1989gs, PhysRevA.59.971, 1601.07772, Rundle2018, Klimov055303, Garon2015}.
For example, there have been various proposals put forward that use a continuous Wigner function to reveal correlations between DV systems~\cite{Garon2015,Mukherjee2018,Rundle2016}.
These methods have further been validated through the direct measurement of phase-space to reveal quantum correlations~\cite{Rundle2016,Leiner2017,Tian2018,Chen2019}. 
Recently this has been extended to experiments validating atomic Schr\"odinger cat states of up to 20 superconducting qubits~\cite{Song2019}.

A case that has not been explored in much detail is the phase-space representation of CV-DV hybridization. 
This hybridisation is seen in many applications of quantum technologies, including simple gate models for quantum computers, such as hybrid two-qubit gates \cite{Reiserer2014,Hacker2019}, and CV microwave pulse control of DV qubits~\cite{HarocheRaimond}.
The generation of hybrid quantum correlations within CV-DV hybrid\footnote{From now on, we shall refer to CV-DV hybrid states as simply `hybrid states', dropping `CV-DV'.} systems commonly takes place within the framework of cavity quantum electrodynamics, that describes the interaction between a two-level quantum system and a single mode of a microwave field.
These models can be further used to describe the effect of circuit quantum electrodynamics, and to consider the interaction of the microwave field with an artificial atom.
Analyzing these interactions within the framework of the Jaynes-Cummings model~\cite{Jaynes1963} allows us to display how quantum information is shared between the CV and DV systems.

Recently there have been theoretical phase-space treatments of hybrid systems~\cite{1601.07772,Rundle2018,Arkhipov2018}, including possible applications to quantum chemistry~\cite{Davies2018}.
Other methods for combining CV Wigner function tomography with other representations of DV systems have been created~\cite{Jeong2014,Vlastakis2015,Sperling2017}.
The visualization technique used in \Ref{Davies2018} displays heterogeneous degrees of freedom, highlighting the power of a hybrid Wigner function approach for visualizing correlations.
This approach also demonstrates how many of the correlations are lost when using standard phase-space methods, such as the reduced Wigner function.
A hybrid phase-space representation, of all the information within these hybrid systems, is crucial for a more complete understanding of CV-DV hybridization, and its physical properties~\cite{Monroe1131,PhysRevLett.77.4887,Deleglise:2008gt}.
This understanding will be especially helpful for advancing quantum technologies~\cite{Andersen2015,Morin2014,PhysRevLett.109.240501,Hacker2019,Gottesman2001}, in particular quantum communication where CV-DV hybridization has been used for teleportation~\cite{Lee2011,Andersen2013,Takeda2013} and entanglement distillation~\cite{VanLoock2006,Ourjoumtsev2007,Datta2012}.

Using the procedure laid out in~\Ref{Rundle2018} to generate any quantum state in phase space, and adapting the visualization method from~\Ref{Davies2018}, we show how the Wigner function of a hybrid system can be intuitively represented.
We begin by presenting examples of important states for CV and DV systems, illustrating how our representation makes correlation information clear.
We extend our analysis using the Jaynes-Cummings model to show how intuitive this representation can be.
The results open new directions for the use of phase-space methods in hybrid quantum systems.

\section{The Wigner function}
The Wigner function is traditionally introduced as the Fourier transform of an autocorrelation function~\cite{Wigner1932, Wigner1984}. 
Here it is more suitable to consider a general Wigner function of some arbitrary operator $\OpA$, defined as~\cite{Groenewold1946}
\begin{equation}\label{TheWignerFunction}
	W_{\OpA}(\Omega) = \Trace{\OpA\,\OpPi(\Omega)},
\end{equation}
where $\OpPi(\Omega)$ is the displaced parity operator for some parameterization of phase space $\Omega$.
The displaced parity operator is defined through displacing a generalized parity operator~\cite{Rundle2018}, and for the CV Wigner function is~\cite{Glauber1969}
\begin{equation}\label{CVKernel}
	\OpPi_f(\alpha) = 2\OpD(\alpha) \HWPi \OpD^\dagger(\alpha),
\end{equation}
where $\HWPi = \sum_{i=0}^\infty (-1)^i\ket{i}\bra{i}$, written here as an operator in the Fock basis, is the usual parity operator that reflects a point through the origin and
\begin{equation}
  \OpD(\alpha) = \exp\left( \alpha \Opad - \alpha^*\Opa \right)
\end{equation}
is the standard CV displacement operator written using the annihilation and creation operators, $\Opa$ and $\Opad$, respectively.
Note that we have introduced the subscript $f$, for `field', to indicate CV systems. 
The displacement operator can be used to define a coherent state~\cite{Glauber1969}
\begin{equation}\label{coherentStates}
	\ket{\beta}_f = \OpD(\beta)\ket{0}_f,
\end{equation}
as the displacement of the vacuum state, $\ket{0}_f$, generating a new coherent state $\ket{\beta}_f$.

As shown in~\Ref{1601.07772,Rundle2018}, a similar approach to~\Eq{CVKernel} can be used to generate Wigner functions for arbitrary quantum systems.
For two-level DV systems, for example, 
\begin{equation}\label{spinKernel}
    \SpinPi(\SpinPar) = \OpU(\SpinPar,\Phi)\SpinPi\OpU^\dag(\SpinPar,\Phi), 
\end{equation}
where the generalized parity, $\SpinPi$, for a single, two-level, system is ${\SpinPi=(\Bid+\sqrt{3}\hat\sigma_z)/2}$~\cite{1601.07772,Rundle2016,Rundle2018}. 
Note that the subscript $a$ here indicates that this is a state for the `atom', or DV system.
The analogue of the displacement operator, $\OpU(\SpinPar,\Phi)$, given in terms of Euler angles, is
\begin{equation}
  {\OpU(\SpinPar,\Phi) = \exp(\ui\Sz\phi) \exp(\ui\Sy\theta) \exp(\ui\Sz\Phi)}
\end{equation}
for the standard Pauli matrices $\Sy$ and $\Sz$. 
Note as the parity operator commutes with $\Sz$, the $\Phi$ term does not contribute, and the DV Wigner function depends only on $\theta$ and $\phi$, allowing it to be plotted on the surface of a sphere.
Note that by DV Wigner function, we mean the Wigner function for DV systems; the Wigner function used here is however parameterized over the continuous variables $\theta$ and $\phi$.

\begin{figure}[bt!]
	\centering
    \includegraphics[width = \linewidth]{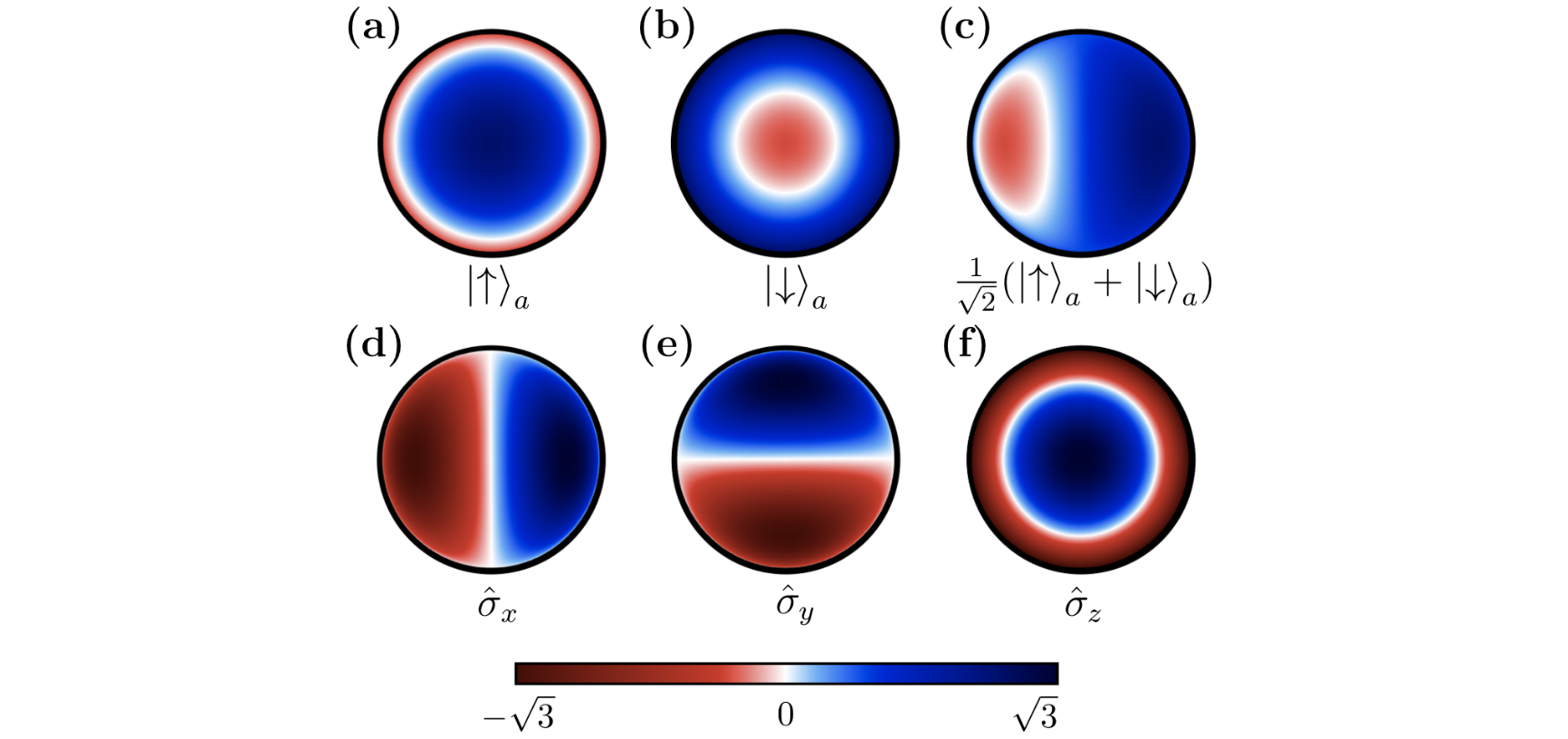}
\caption{\label{FIG:SPINS} 
    Shown here are six example qubit Wigner functions using the Lambert azimuthal equal-area projection, that maps a sphere onto a circle where the north pole is mapped to the centre and the south pole is on the perimeter.
    Three single-qubit pure states are shown in~\SubTo{a}{c}, where \Sub{a} and~\Sub{b} are the eigenstates of $Sz$, $\Excited_a$ and $\Ground_a$, with eigenvalues $\pm 1$ respectively. 
    \Sub{c} is the equal superposition of the states in \Sub{a} and~\Sub{b}, $(\Excited_a+ \Ground_a)/\sqrt{2}$.
    \SubTo{d}{f} show the qubit Wigner functions of the three Pauli matrices, $\Sx$, $\Sy$, and $\Sz$ respectively.
}
\end{figure}

Figure \ref{FIG:SPINS} shows examples of the DV Wigner function generated by~\Eq{spinKernel} for some simple qubit states.
Each of the DV Wigner functions presented in~\Fig{FIG:SPINS} is plotted following~\Ref{PhysRevA.85.022113}, using the Lambert azimuthal equal-area projection~\cite{Lambert}.
This projection is area preserving and maps the surface of a sphere to polar coordinates, with the north pole mapped to the centre of the disc and the south pole to the outer boundary. 
The equator of the sphere is projected onto a concentric circle, with a radius $1/\sqrt{2}$ times the radius of the entire circle, this is explicitly seen as the white circle in~\FigsSub{FIG:SPINS}{f}.
This means that the Lambert azimuthal equal-area projection allows us to view the entire surface of the sphere as a circle.
The reason for using this area-preserving mapping, rather than an angle-preserving mapping, is because we are dealing with a probability distribution function.
By definition, the integral over a volume determines the probability; area-preserving therefore translates into probability-preserving.
A consequence of this mapping is that in some regions of phase space, the quasi-probability distribution appears warped.
For instance, the first three states in \FigsSub{FIG:SPINS}{a}~-~\Sub{c} are all rotations of one another on a sphere.

The DV Wigner functions presented in~\FigsSub{FIG:SPINS}{a}~-~\Sub{c} are standard two-level quantum states, where \FigsSub{FIG:SPINS}{a} and~\Sub{b} are the $\pm1$ eigenstates of the $\Sz$ operator, $\Excited_a$ and $\Ground_a$ respectively.
The state in~\FigSub{FIG:SPINS}{c} is the equal superposition of $\Excited_a$ and $\Ground_a$, or the positive eigenstate of $\Sx$.
In all the presented states, there are negative values in the DV Wigner function.
Importantly, in the DV Wigner function for qubits, negative volume, as well as being an indicator of non-classicality, is also a measure of purity~\cite{Arkhipov2018}.
This is because discrete system coherent states are fundamentally quantum; regardless of whether the system is the polarization of a photon or the direction of spin in an electron.

More generally, in both CV and DV Wigner functions, negative values arise as a consequence of self-interference.
In the CV Wigner function this arises from non-Gaussianity~\cite{HUDSON}, and can be seen in the Fock states (excluding the vacuum state) or in superpositions of Gaussian states, see~\Fig{clusterFocks} for an example, discussed later in the paper.
This explains why negative values have been used as a measure of quantumness, however there is one notable exception, the non-negative, entangled, Gaussian CV two-mode squeezed state.

Since the Gaussian states of a DV Wigner function can be visualized on a sphere, the emergence of self-interference is now inevitable, due to the inherent geometry of the sphere.
For example, the Wigner function for the state $\Excited_a$ has a Gaussian distribution centred at the north pole; as this Gaussian distribution tends towards zero, near the south pole, there is an emergence of negative quasi-probabilities. 
This negativity in the Wigner function is manifested as a result of self-interference, as the quantum coherences interfere with each other at the south pole.
As the number of levels is increased (from the two-level system) in the DV Wigner function and take the infinite limit\footnote{The general Wigner function for any system in the displaced parity formalism can be found in~\Ref{Rundle2018}.}, the \SU{2} DV Wigner function tends towards the Heisenberg-Weyl group, returning to the standard CV Wigner function.
This is because the effective size of the sphere increases, decreasing the relative size of the Gaussian.
In the infinite limit, the negativity in the Wigner function is completely eliminated, since the Gaussian can no longer interact with itself on the opposite side of the sphere.

Although the example states so far have been density operators for pure states, the general formalism in \Eq{TheWignerFunction} allows for the Wigner function to be generated for any arbitrary operator.
To emphasize this, in~\FigsSub{FIG:SPINS}{d}~-~\Sub{f} are the DV Wigner representation of each of the three Pauli operators. 
In general, Wigner function exhibit the normalization condition
\begin{equation}
  \int_\Omega \ud\Omega \; W_{\OpA}(\Omega) = \Trace{\OpA}.
\end{equation}
For normal density operators, this yields unity, as would be expected for any probability distribution function.
For the Pauli operators however, $\Trace{\hat{\sigma}_i} = 0$, where $i~=~\{x,y,z\}$, therefore $\int_\Omega \ud\Omega \, W_{\hat{\sigma}_i}(\Omega) = 0$.
The tracelessness of these matrices can be seen in~\Fig{FIG:SPINS}~\SubTo{d}{f} by noting that the negative and positive volumes are equivalent and therefore cancel.
This feature will be key to several of our observations later in this work.

For a CV-DV hybrid system, the total displaced parity operator is simply the tensor product of the displaced parity operator for each subsystem~\cite{1601.07772, Rundle2018,Rundle2016}
\begin{equation}\label{fullWF}
	\OpPi(\HWPar,\SpinPar) = \HWPi(\HWPar)\otimes\SpinPi(\SpinPar), 
\end{equation}
yielding a hybrid Wigner function for a density matrix $\DO$
\begin{equation}\label{full_WF}
	W_{\DO}(\HWPar,\SpinPar) = \Trace{\DO \; \OpPi(\HWPar,\SpinPar)}.
\end{equation}

\begin{figure}
	\centering
	\includegraphics[width = \linewidth]{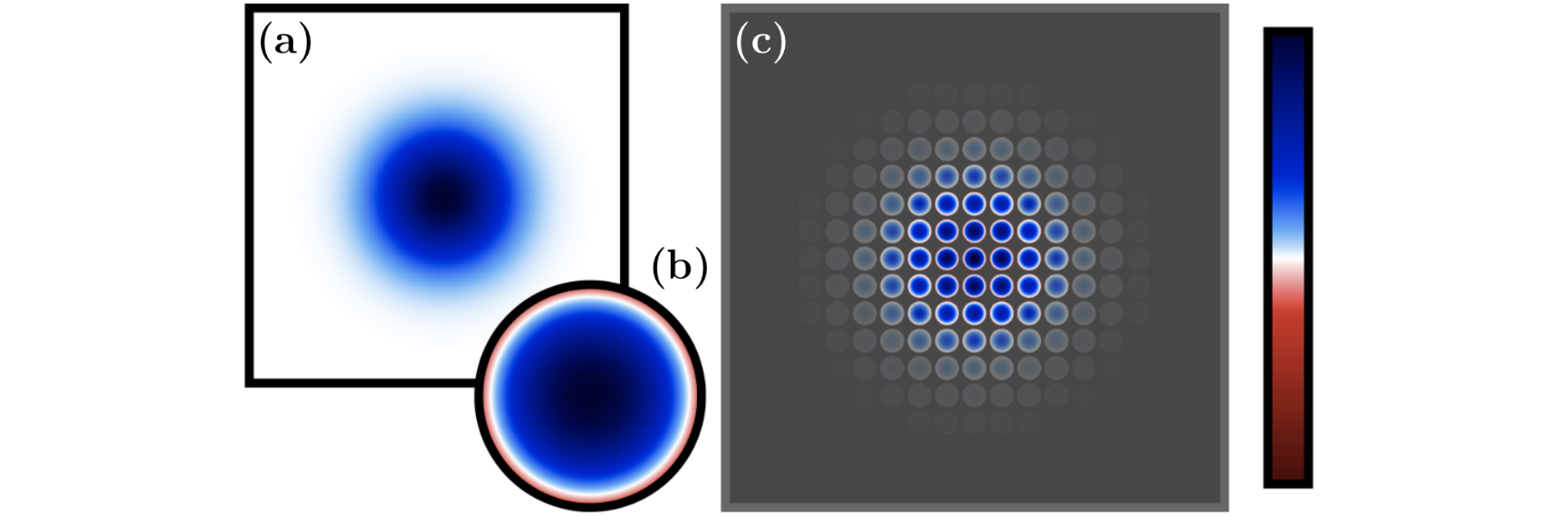}
	\caption{\label{exampleQSS} Example Wigner function for the product of the CV vacuum state and a DV excited state, $\ket{0}_f\Excited_a$, where \Sub{a} and \Sub{b} show the reduced Wigner functions for the continuous-variable (CV) and discrete-variable (DV) Wigner functions respectively.
	In~\Sub{c} is the full Wigner function of the hybrid system, where the CV phase space is split up as a discrete grid. 
	At each of these discrete points the DV Wigner function at that point in phase space is plotted.
	The transparency of each of the DV Wigner functions is proportional to the maximum quasi-probability at that point in CV phase space.
	The colour bar is white at $0$ with limits $\pm 2$ for \Sub{a}, $\pm (1+\sqrt{3})/2$ for \Sub{b}, and $\pm (1+\sqrt{3})$ for \Sub{c}.}
\end{figure}
Hybrid systems generated with~\Eq{full_WF} usually have more degrees of freedom than is convenient to plot. 
For this reason, many approaches that use phase-space methods to treat hybrid systems use reduced Wigner functions, rather than considering the full phase space of the composite system.
To give a full picture of the quantum correlations found between the two systems, a method similar to that introduced in~\Ref{Davies2018} can be used.
As an example of the utility of this method, the fully separable state, $\ket{\beta}_f\Excited_a$, is shown in~\Fig{exampleQSS}.
The reduced Wigner functions for CV and DV degrees of freedom are presented in~\FigsSub{exampleQSS}{a} and~\Sub{b} respectively.
In~\FigSub{exampleQSS}{c} we apply the method first presented in~\Ref{Davies2018} to plot the phase-space representation of this state.

Specifically,~\FigSub{exampleQSS}{c} was created by first dividing the CV phase space into discrete points on a rectangular map.
Each of these discrete points is then associated with a discrete complex value $\alpha$, equally spaced across the phase space grid.
For each set point $\alpha$, the values of the Wigner function for $\theta$ and $\phi$ degrees of freedom are calculated, with the Wigner function at that point plotted using the Lambert projection.
This produces a DV Wigner function at each $\alpha$ in CV phase space.
The transparency of each disc is then set proportional to the absolute maximal value of the phase space at that point, $ \max_{\theta,\phi}|W_{\DO}(\alpha,\theta,\phi)|$.
For example, to generate the disc at the centre of~\FigSub{exampleQSS}{c}, we calculate $W_{\DO}(\alpha = 0 ,\theta,\phi)$, resulting in a DV Wigner function for $\Excited_a$, and then modify the amplitudes of the quasi-probabilities using the value of $\alpha$.
This is then repeated for every $\alpha$.
As a result,~\FigSub{exampleQSS}{c} has the same form as a coherent state, dictated by the CV Wigner function, with every point in phase space having an $\Excited_a$ DV Wigner function.
The difference in this method, in comparison to~\Ref{Davies2018}, is that here the transparency is not set by integrating out the qubit degrees of freedom; such an approach leads to a loss of quantum correlations in the systems of interest.

\section{Visualizing correlations in hybrid quantum systems}\label{VisualisingQuantumCorrelations}
Quantifying different types of correlations in quantum systems is a key area of research that has received a great deal of attention~\cite{Bennett1996,Vedral1997,Wootters1998,Vedral1998,PhysRevA.23.236,Eisert1999,Henderson2001,Adesso2016}.
In parallel, phase-space methods have been utilized as a tool to identify and categorize quantum correlations~\cite{Wallentowitz1997,Agudelo2017,Sperling2017,Sperling2018,Sundar2019}.
Further, these methods have been used to generate measures based on the emergence of negative quasi-probabilities in the Wigner function~\cite{Kenfack2004,Arkhipov2018,Taghiabadi2016,Siyouri2016}.
However, due to the higher number of degrees of freedom, visually representing correlations in composite systems is more difficult.
We now show how our technique produces definite signatures of both quantum and classical correlations, that can be discerned for hybrid quantum systems.
When dealing with quantum information processing with two coupled qubits, the distinction between these two types of correlations is important.
Beginning with how correlations that arise from superposition appear, we will describe our choices of DV and CV qubits and how the encoding of quantum information is represented on these qubits.

Certain similarities are seen between DV and CV systems, whether in structure, choice in qubit, or in appearance of the quantum correlations that manifest.
These similarities will be demonstrated here, by showing how quantum information can be encoded onto different types of state.
Encoding quantum information onto quantum states can be done in various ways, including a variety of approaches even within the same system~\cite{Andersen2015}.
We will therefore begin by using the simplest case of a DV qubit for quantum information processing.
Since the DV systems used here are two-level systems, the encoding of quantum information is straightforward; a bit value $0$ or $1$ is simply assigned to each of the two levels, $\Excited_a$ and $\Ground_a$ respectively.
The DV $0$ bit is now represented visually by~\FigSub{FIG:SPINS}{a}, likewise the $1$ bit value is represented by~\FigSub{FIG:SPINS}{b}.
Furthermore, a general pure superposition state
\begin{equation}\label{DVQubitState}
	a_a\Excited_a + b_a\Ground_a,
\end{equation}
where $|a_a|^2+|b_a|^2=1$, allowing any weighted superposition between  $0$ and $1$. 
When $a_a=b_a=1/\sqrt{2}$, an equal superposition is yielded and is represented visually by~\FigSub{FIG:SPINS}{c}.

This binary choice becomes more complicated when assigning bit values to a CV qubit.
Although, there are various ways to encode quantum information onto a CV system creating similarities between CV and DV systems.
Since the Hilbert space is infinite, there are different constraints on assigning qubit values.
We will now demonstrate two examples of CV qubits, comparing the results with the DV qubits

\subsection{Fock state qubits}
Fock states are orthogonal and therefore a natural choice for quantum information processing.
For simplicity we consider the vacuum and one-photon Fock states, $\ket{0}_f$ and $\ket{1}_f$ respectively. 
We can now form the analogy with the DV qubit state by assigning bit values to these states $0\rightarrow\ket{0}_f$ and $1\rightarrow\ket{1}_f$.

\begin{figure}
	\centering
	\includegraphics[width = \linewidth]{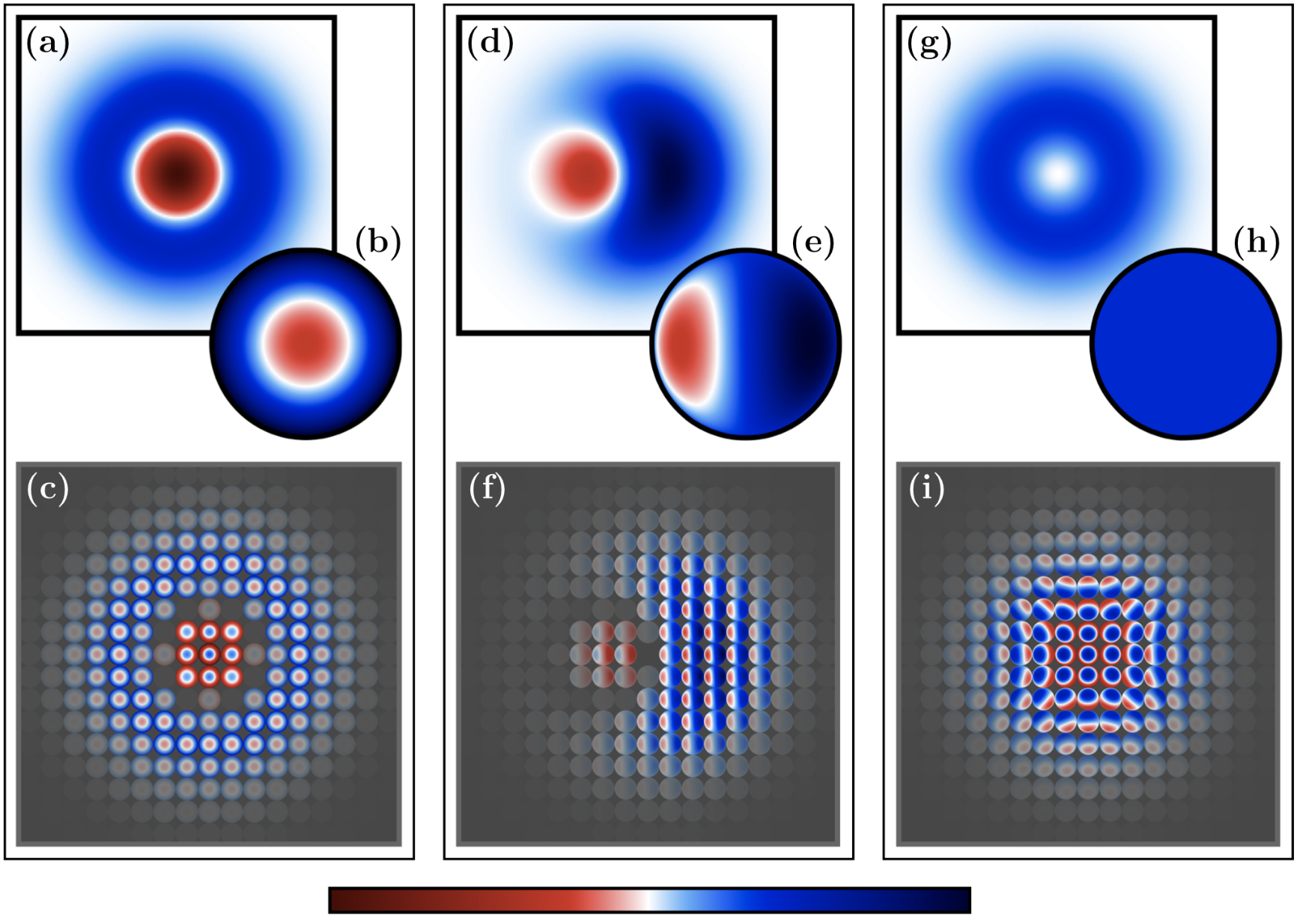}
	\caption{\label{clusterFocks} Examples of Fock states coupled to DV qubits. 
	\Sub{a}~-~\Sub{c} show the state $\ket{1}_f\Ground_a$. 
	\Sub{d}~-~\Sub{f} are the state $(\ket{0}_f+\ket{1}_f)(\Excited_a+\Ground_a)/2$. \Sub{g}~-~\Sub{i} are the entangled state $(\ket{0}_f\Excited_a+\ket{1}_f\Ground_a)/2$.
	\Sub{a},\Sub{d}, and \Sub{g} show the reduced CV Wigner functions,
	\Sub{b},\Sub{e}, and \Sub{h} are the reduced DV Wigner functions and  \Sub{c},\Sub{f}, and \Sub{i} are the full hybrid Wigner functions.
	The colour bar is white at $0$ with limits $\pm 2$ for the reduced CV Wigner function, $\pm (1+\sqrt{3})/2$ for reduced DV Wigner function, and $\pm (1+\sqrt{3})$ for hybrid Wigner function.}
\end{figure}

Comparison of the Wigner functions for the DV and the CV Fock qubits can be found in \FigsSub{exampleQSS}{a} and~\Sub{b}; where in the Lambert projection, the DV qubit in \FigSub{exampleQSS}{b} has a similar Gaussian form as the vacuum state in \FigSub{exampleQSS}{a}.
In fact, the DV qubit basis states are discrete analogues of the Fock states.
Therefore, the presence of the negative values in the DV qubit states becomes more apparent by considering the one-photon Fock state $\ket{1}_f$ and the DV qubit state $\Ground_a$ (in \FigsSub{clusterFocks}{a} and~\Sub{b} respectively).
The orientation of the DV qubit is somewhat arbitrary, the $\Excited_a$ and $\Ground_a$ states are orthogonal rotations of one another; therefore, the DV qubit states share properties of both the $\ket{0}_f$ and $\ket{1}_f$ Fock states.
This analogy can be seen further in~\FigsSub{clusterFocks}{d} and~\Sub{e}, where the Wigner functions for the states $(\ket{0}_f+\ket{1}_f)/\sqrt{2}$ and $(\Excited_a+\Ground_a)/\sqrt{2}$ are shown respectively.

Also in \Fig{clusterFocks}, we show the hybrid Wigner functions for these states.
In \FigSub{clusterFocks}{c} we show the product of \FigsSub{clusterFocks}{a} and~\Sub{b}.
The product of \FigsSub{clusterFocks}{d} and~\Sub{e} is shown in \FigSub{clusterFocks}{f}.
Since in both cases the CV and DV qubits are separable, the pattern of the hybrid phase space is similar to that found in \Fig{exampleQSS}.
The separability is evident by the existence of a DV Wigner function at every point in CV phase space, with the amplitude modulated by the CV Wigner function at that point.
For both of the hybrid Wigner functions in \FigsSub{clusterFocks}{c} and~\Sub{f}, the negative regions in the CV Wigner functions affect the sign of the DV Wigner function, inverting the positive and negative quasi-probabilities at those points in CV phase space.

Having established that the hybrid Wigner function allows local correlations to be discerned reliably, we now demonstrate how quantum correlations arising between subsystems in this type of hybrid system manifest.
Entanglement in Fock hybrid states, a Bell-Fock state\footnote{Bell state for an entangled DV qubit with a CV Fock qubit.}, $(\ket{0}_f\Excited_a+\ket{1}_f\Ground_a)/\sqrt{2}$, is shown in~\FigSub{clusterFocks}{i}.
The full Wigner functions for bipartite Bell-Fock states have a distinctive pattern, reminiscent of the spin-orbit coupled state from~\Ref{Davies2018}, where there is a twisting of the DV Wigner functions dependent on the point in CV phase space.
This DV dependence on the CV Wigner function is indicative that there is coupling between the two subsystems. 
This is a signature one should look for when investigating quantum correlations in this type of hybrid state.

Comparing the hybrid Wigner function in \FigSub{clusterFocks}{i} to the reduced Wigner function for the CV and DV qubits in \FigsSub{clusterFocks}{g} and~\Sub{h} respectively, we see the importance in considering the full phase space for entangled states such as this.
It can be seen in \FigsSub{clusterFocks}{g} and~\Sub{h} how correlations between the two systems are lost when considering the reduced Wigner functions, leaving only statistical mixtures of the basis states in each case.

\subsection{Coherent state qubits}
Another choice in creating a CV qubit is to encode quantum information onto coherent states~\cite{Ralph2003,Gilchrist2004}.
Unlike with the Fock CV qubit, the coherent state basis is an overcomplete basis where there is some degree of overlap between any two coherent states.
However with sufficient distance between two coherent states, this overlap is negligible.
For simplicity, our example states will be real values of $\beta$, where the two levels are set to the values $\beta_1 = -\beta_2 = \beta$.

\begin{figure}
	\centering
    \includegraphics[width = \linewidth]{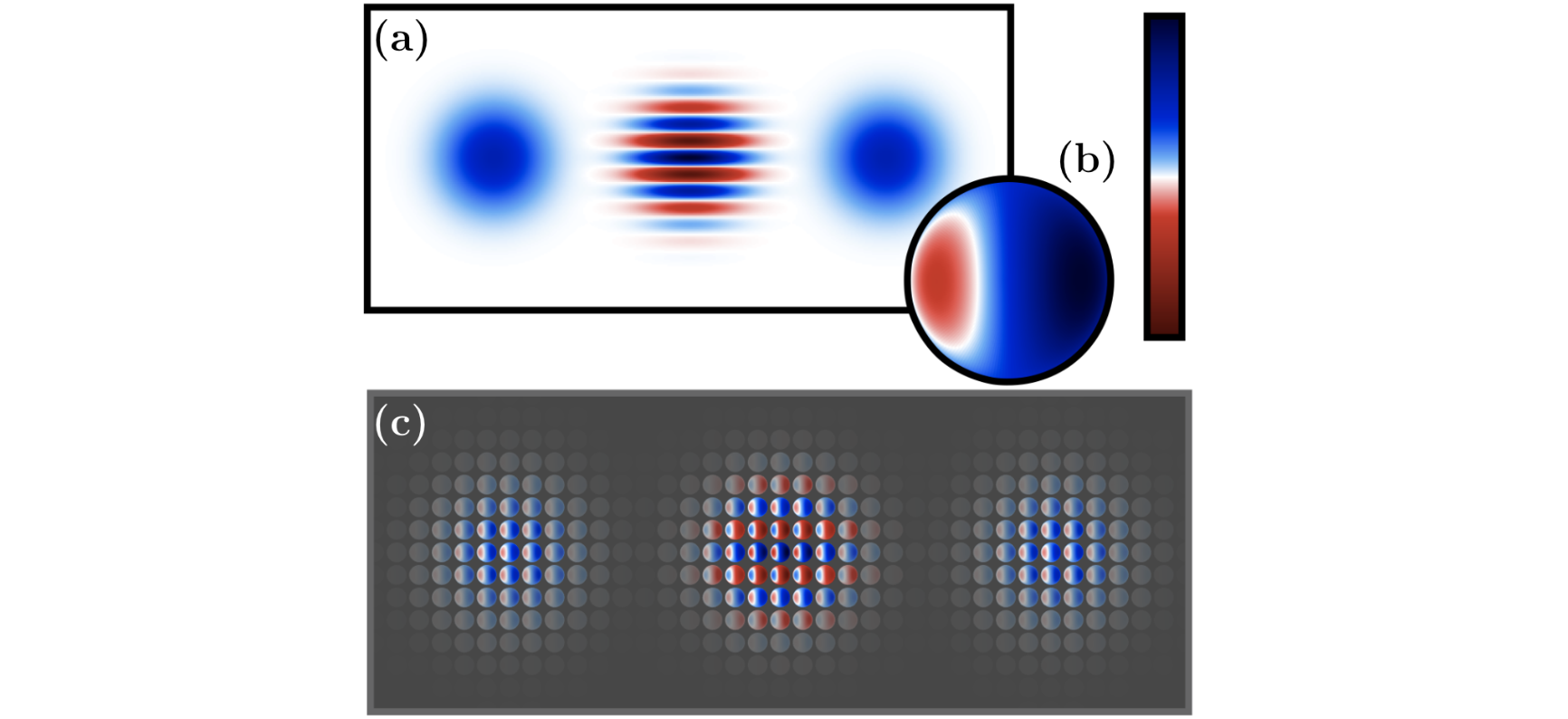}
    \caption{\label{ExampleSeparableCats}
    Here is an example of the Wigner representation of a Schr\"odinger cat state coupled to a qubit, $(\ket{\beta}_f+\ket{-\beta}_f)(\Excited_a+\Ground_a)/2$, where $\ket{\beta}$ is a coherent state centred at $\beta$ for $\beta = 3$.
    \Sub{a} shows the reduced CV Wigner function and \Sub{b} shows the reduced DV Wigner function. 
    \Sub{c} shows the hybrid Wigner function.
    The colour bar is white at $0$ with limits $\pm 2$ for \Sub{a}, $\pm (1+\sqrt{3})/2$ for \Sub{b}, and $\pm (1+\sqrt{3})$ for \Sub{c}.}
\end{figure}

We then label each of the coherent states as a certain bit value; for instance $0\to\beta_1$ and $1\to\beta_2$.
This creates a qubit in the form of a Schr\"odinger cat state~\cite{Gilchrist2004}, with the general qubit state being
\begin{equation}\label{CVQubitState}
	a_f\ket{\beta}_f + b_f\ket{-\beta}_f,
\end{equation}
as in~\Eq{DVQubitState}.
This means that there is a coherent state at $\beta$ when $a_f=1$ and a coherent state at $-\beta$ when $b_f=1$.
The superposition state $a_f = b_f = 1/\sqrt{2}$ produces the Schr\"odinger cat state shown (for $\beta=3$) in~\FigSub{ExampleSeparableCats}{a}.

Coupling the CV and DV qubits in~\FigsSub{ExampleSeparableCats}{a} and~\Sub{b} generates the full Wigner function in~\FigSub{ExampleSeparableCats}{c}.
Explicitly, this is the state
\begin{equation}\label{SepCatAtom}
	\frac{1}{2}\!\left(\ket{\beta}_f+\ket{-\beta}_f\right)\left(\Excited_a+\Ground_a\right).
\end{equation}
Since the full system is a simple tensor product of the two qubits, the subsystems are separable.
The separability between these states is seen in the full Wigner function in~\FigSub{ExampleSeparableCats}{c}.
The image of the CV Schr\"odinger cat state is visible as a discrete grid, with the DV Wigner function for the state at every point.

\begin{figure}
	\centering
    \includegraphics[width = \linewidth]{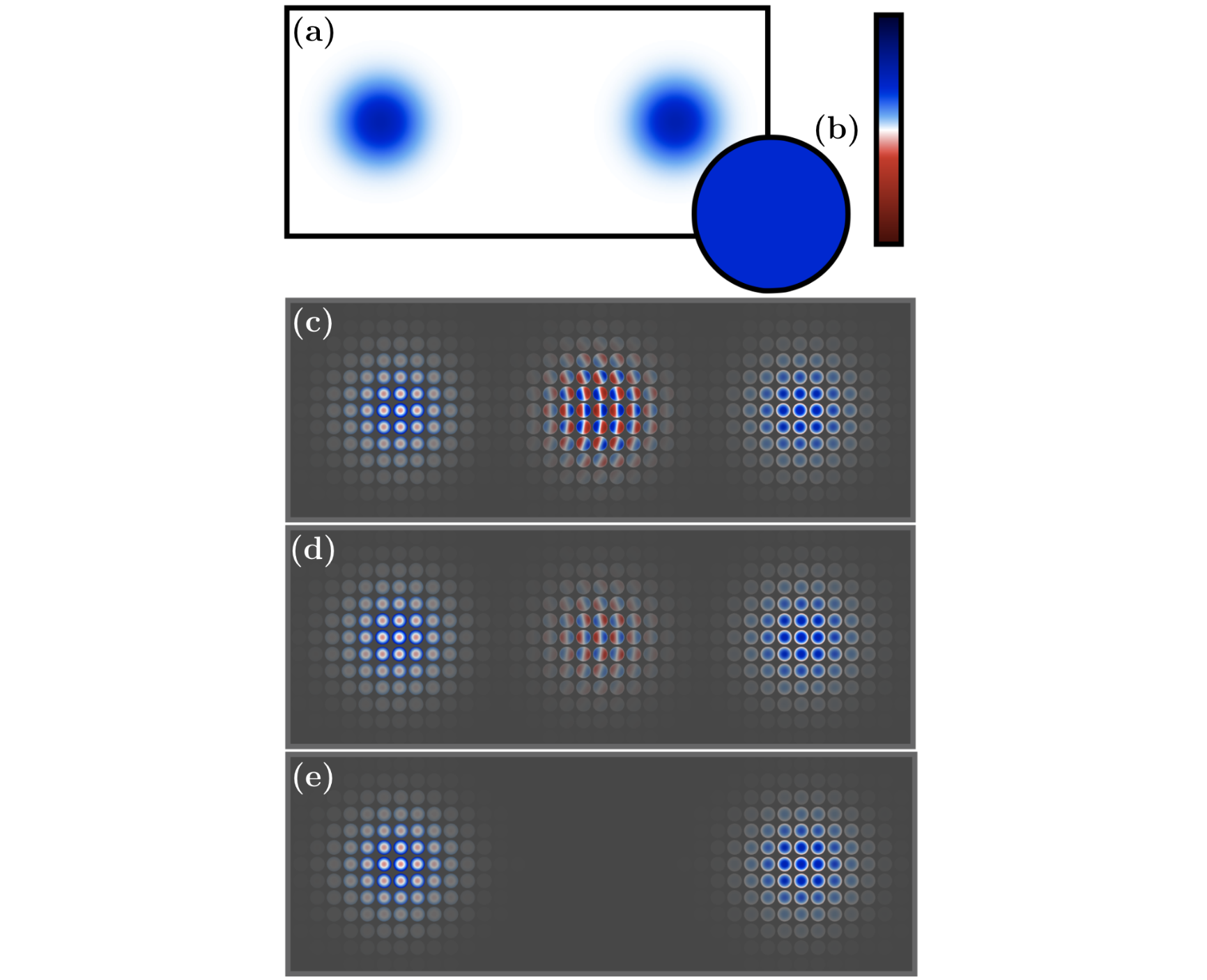}
\caption{\label{ExampleEntangledCats} Here are examples of the Wigner representation of a lossy entangled Bell-cat state, with varying values of loss.
\Sub{a} shows the reduced CV Wigner function and \Sub{b} shows the reduced DV Wigner function. 
The reduced Wigner functions remain the same for the following three example states.
\Sub{c} shows the full Wigner function for the state with no loss $(\ket{\alpha}_f\Excited_a+\ket{-\alpha}_f\Ground_a)/\sqrt{2}$.
\Sub{d} shows partial loss of the quantum correlations.
\Sub{e} shows a fully mixed version of the state $(\Proj{\beta}{\beta}{f}\Proj{e}{e}{a} + \Proj{-\beta}{-\beta}{f}\Proj{g}{g}{a})/2$.
The colour bar is white at $0$ with limits $\pm 2$ for \Sub{a}, $\pm (1+\sqrt{3})/2$ for \Sub{b}, and $\pm (1+\sqrt{3})$ for \Sub{c}.}
\end{figure}

Given the difference in the local correlations between the two choices of CV qubit, it is now worthwhile to demonstrate how the signature of the non-local correlations differ for the coherent state CV qubits.
The hybrid analogue of a Bell state for coherent states, the Bell-cat state, is
\begin{equation}\label{BellCatState}
	\frac{1}{\sqrt{2}}\left(\ket{\beta}_f\Excited_a + \ket{-\beta}_f\Ground_a \right).
\end{equation}
Since many of the correlations in this state are due to entanglement, the standard approach of using reduced Wigner functions is insufficient, as seen in~\FigsSub{ExampleEntangledCats}{a} and~\Sub{b}.
Neither reduced Wigner function has visible quantum correlations, yielding two mixed states.
This issue motivated other approaches to tomography and state verification for such states, for instance~\Ref{Vlastakis2015} used reduced CV Wigner functions in different Pauli bases to show Bell's inequality.
Other tomography methods for entangled hybrid systems, such as~\Ref{PhysRevLett.109.240501}, also take into consideration the problems of a reduced phase-space representation of a hybrid entangled state.
Although approaches such as these give a better appreciation of the quantum correlations, they still only provide glimpses of the nature of the full quantum state.

The hybrid Wigner function for~\Eq{BellCatState} is shown in~\FigSub{ExampleEntangledCats}{c}.
Comparing our representation with the reduced Wigner function treatment, the quantum correlations are now visible, manifesting as interference terms between the two coherent states.
The nature of these quantum correlations is completely lost when the full Wigner function is not generated.
Further, within the quantum correlations, the qubit states approach traceless states, as in~\FigsSub{FIG:SPINS}{d} -~\Sub{f}, where the state at the very centre, $\alpha = 0$, is in fact the $\Sx$ Pauli matrix.
It is important to note at this point that the manifestation of traceless here, found only in the hybrid phase-space picture, are a signature of quantum correlations.
Some existing tomography methods can pick up these correlations, however their full nature is not captured. 
For example, measuring the reduced Wigner functions results in a loss of quantum and classical correlations, as demonstrated in~\FigsSub{ExampleEntangledCats}{a} and~\Sub{b}.
This makes classical and quantum correlations, for this kind of state, indistinguishable.
The ability to obtain signatures to distinguish between classical and quantum correlations is important in determining the suitability of states in quantum information processing.

To highlight this, we now consider two further examples of states that have the same reduced CV and DV Wigner functions.
Though the degree of quantum correlations differ for each state.
The general state is 
\begin{equation}\label{EntangledState}
	\frac{1}{2}\left(\Proj{\beta}{\beta}{f}\Proj{e}{e}{a} + \eta \Proj{\beta}{-\beta}{f}\Proj{e}{g}{a} + \eta \Proj{-\beta}{\beta}{f}\Proj{g}{e}{a} +  \Proj{-\beta}{-\beta}{f}\Proj{g}{g}{a} \right),
\end{equation}
where $\eta$ determines the purity of the state.
When $\eta=1$ \Eq{EntangledState} reduces to~\Eq{BellCatState}.
Changing the value of the loss to $\eta = 0.5$ and then to $\eta = 0$,~\FigsSub{ExampleEntangledCats}{d} and~\Sub{e} are, respectively, generated.
In both, it is clear that the quantum correlations are slowly lost.
The loss of quantum correlations means these states are less useful for quantum information purposes, and analyzing the reduced Wigner functions, unlike our approach, does not provide any insight to this loss.
By using our method to represent the full Wigner function, it is not only possible to distinguish the strength of the quantum correlations but, the signature of classical correlations is revealed.

In~\FigSub{ExampleEntangledCats}{e} is the state
\begin{equation}
  \Half(\Proj{\beta}{\beta}{f}\Proj{e}{e}{a} + \Proj{-\beta}{-\beta}{f}\Proj{g}{g}{a})
\end{equation}
that describes the equal classical probability of finding an excited state at $\beta$ and a ground state at $-\beta$.
The classical correlations that correspond to this probability is shown in our full picture of the Wigner function, where the $\ket{\beta}_f$ coherent state is correlated with $\Excited_a$ states, likewise the $\ket{-\beta}_f$ coherent state is correlated with $\Ground_a$ states.
This process not only reveals that this is the signature of classical correlations, it verifies the case that the traceless states between the two states are a result of the quantum correlations within the hybrid system.

\section{The Jaynes-Cummings model}
Light-matter interaction in the form of quantum electrodynamics (QED) has been an experimental cornerstone in understanding quantum effects.
It has also given a helping hand in the development of quantum information applications, such as single-photon quantum non-demolition measurements acting as two-qubit gates between microwaves and atoms~\cite{HarocheRaimond}.
The standard example of a QED interaction between a two-level DV system and a CV field is the Jaynes-Cummings model~\cite{Jaynes1963}.
Jaynes-Cummings type interactions are the basis for the generation of non-Gaussian states and are well known for showing the collapse and revival of Rabi oscillations~\cite{PhysRev.140.A1051, PhysRevLett.44.1323,PhysRevA.23.236} throughout its evolution.
During this evolution, quantum information is transferred back and forth between the CV and DV systems; through this process, quantum information can then 
manifest as a Schr\"odinger cat state or generate Bell pairs of the sort shown in \FigSub{clusterFocks}{i}.
By using our methods, the transfer of quantum information can be visualized as is swaps between the microwave field and the atom.

The interaction picture of the Jaynes-Cummings model 
\begin{equation}\label{JCM}
    \OpH_{\mathrm{JC}} = \omega (\Opad\OpSm + \Opa\OpSp),
\end{equation}
will be used, where $\omega$ is the field-qubit coupling constant, and the operators $\hat{\sigma}_{\pm} = (\Sx\pm\ui\Sy)/2$ are the qubit raising and lowering operators that transition the state between eigenstates of $\Sz$.

\begin{figure}
	\centering
	\includegraphics[width = \linewidth]{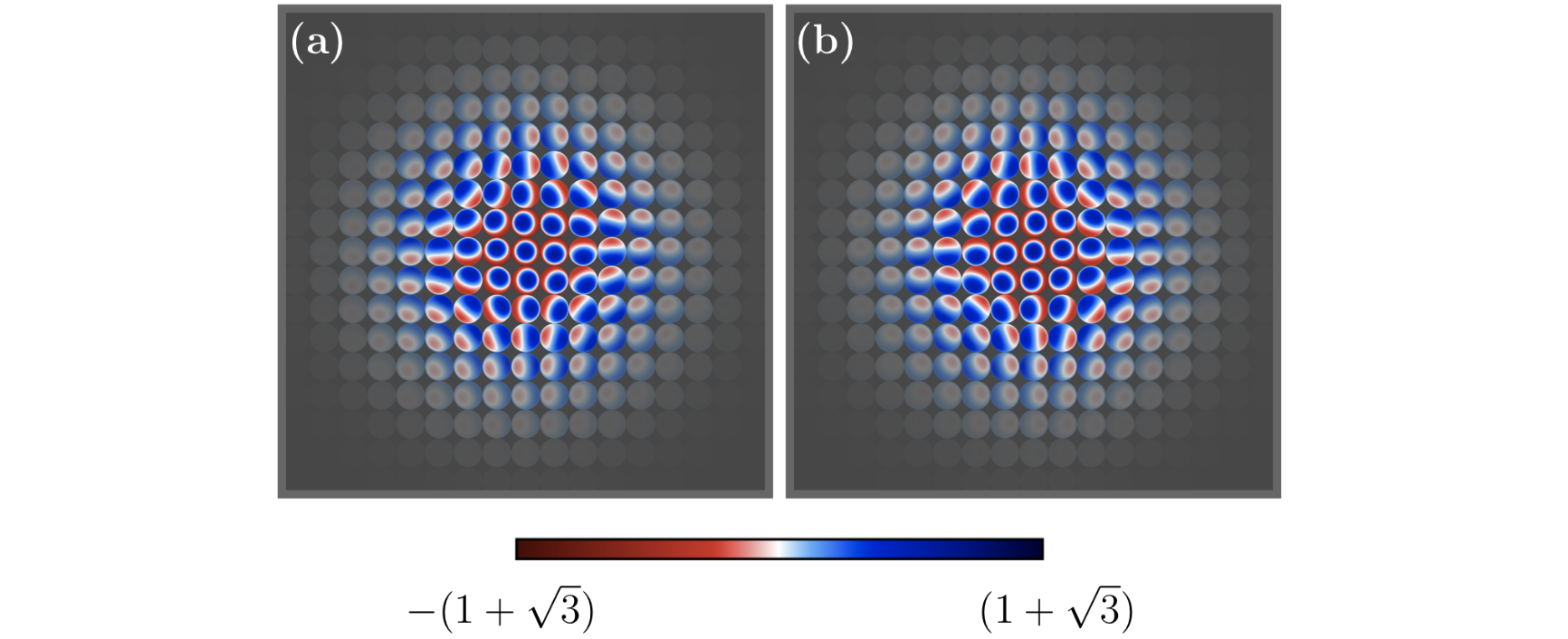}
	\caption{\label{JCFocks} Here we show the Wigner functions for two points in the evolution of the Jaynes-Cummings model with initial state $\ket{0}_f\Excited_a$.
	During the evolution of the Jaynes-Cummings model with this excited state, two entangled Bell-Fock states are generates before returning to the initial state again.
	The two entangled Fock-States are shown here, where the first one in \Sub{a} is the state $(\ket{0}_f\Excited_a - \ui\ket{1}_f\Ground_a)/\sqrt{2}$.
	The second entangled Bell-Fock state in the evolution is shown in \Sub{b}, where the state is $(\ket{0}_f\Excited_a + \ui\ket{1}_f\Ground_a)/\sqrt{2}$.
	The signature of entanglement in these states can be seen in the DV Wigner function dependence on the value of the CV Wigner function, similarly to the example state in \FigSub{clusterFocks}{i}.} 
\end{figure}

Following the example given in~\Sec{VisualisingQuantumCorrelations}, we consider a Fock state basis to model the Jaynes-Cumming model.
Choosing the initial state in the field to be a vacuum state and coupling it to an excited DV qubit results in an evolution that fluctuates between $\ket{0}_f\Excited_a$ and $\ket{1}_f\Ground_a$~\cite{HarocheRaimond}, as shown in~\Fig{exampleQSS} and~\FigsSub{clusterFocks}{d} -~\Sub{g} respectively. 
This means that the evolution can be fully described with the two levels of the Fock state qubit and the DV qubit, allowing us to consider this as an exchange between two qubits. 

The fluctuation as part of this model results in the system continuously transferring quantum information between the two qubits, where the state at time $t$ is 
\begin{equation}
	\ket{\Psi(t)} = \cos(\omega t)\ket{0}_f\Excited_a - \ui\sin(\omega t)\ket{1}_f\Ground_a,
\end{equation}
returning to the initial state at $t = \pi/\omega$.
A video of this evolution is given in supplementary material.
As the information transfers between these two states, throughout one period, two entangled Bell-Fock states are generated
\begin{equation}
	\ket{\Phi^{\pm}}=\frac{1}{\sqrt{2}}\left(\ket{0}_f\Excited_a \pm \ui \ket{1}_f\Ground_a\right),
\end{equation} 
where the full Wigner functions for these states are shown in~\Fig{JCFocks}.
Both of these states have the same reduced Wigner functions, which are not shown here since all Bell-Fock states have the same reduced Wigner functions, shown in~\FigsSub{clusterFocks}{g} and~\Sub{h}.

During Jaynes-Cummings evolution, the first of the Bell-Fock states appears at $t =\omega^{-1}\pi/4$, where the state $\ket{\Psi(\omega^{-1}\pi/4)} = \ket{\Phi^{-}}$. 
This first Bell-Fock state is shown in~\FigSub{JCFocks}{a}, the second $\ket{\Psi(3\omega^{-1}\pi/4)} = \ket{\Phi^{+}}$ is given in~\FigSub{JCFocks}{b}.
Comparing these two states to~\FigSub{clusterFocks}{i}, even though the reduced Wigner functions are identical, the difference the phase plays in the full hybrid Wigner functions is apparent.
Extrapolating to another choice of phase, for example $(\ket{0}_f\Excited_a - \ket{1}_f\Ground_a)/\sqrt{2}$, the full hybrid Wigner function is similar to~\FigSub{clusterFocks}{i} with each of the DV Wigner functions pointing in the orthogonal directions.
The quantum correlations that arise in this form of hybrid system have a unique signature which can best be described as a twisting of the DV Wigner function at points in CV phase space.

\begin{figure}
    \includegraphics[width = \linewidth]{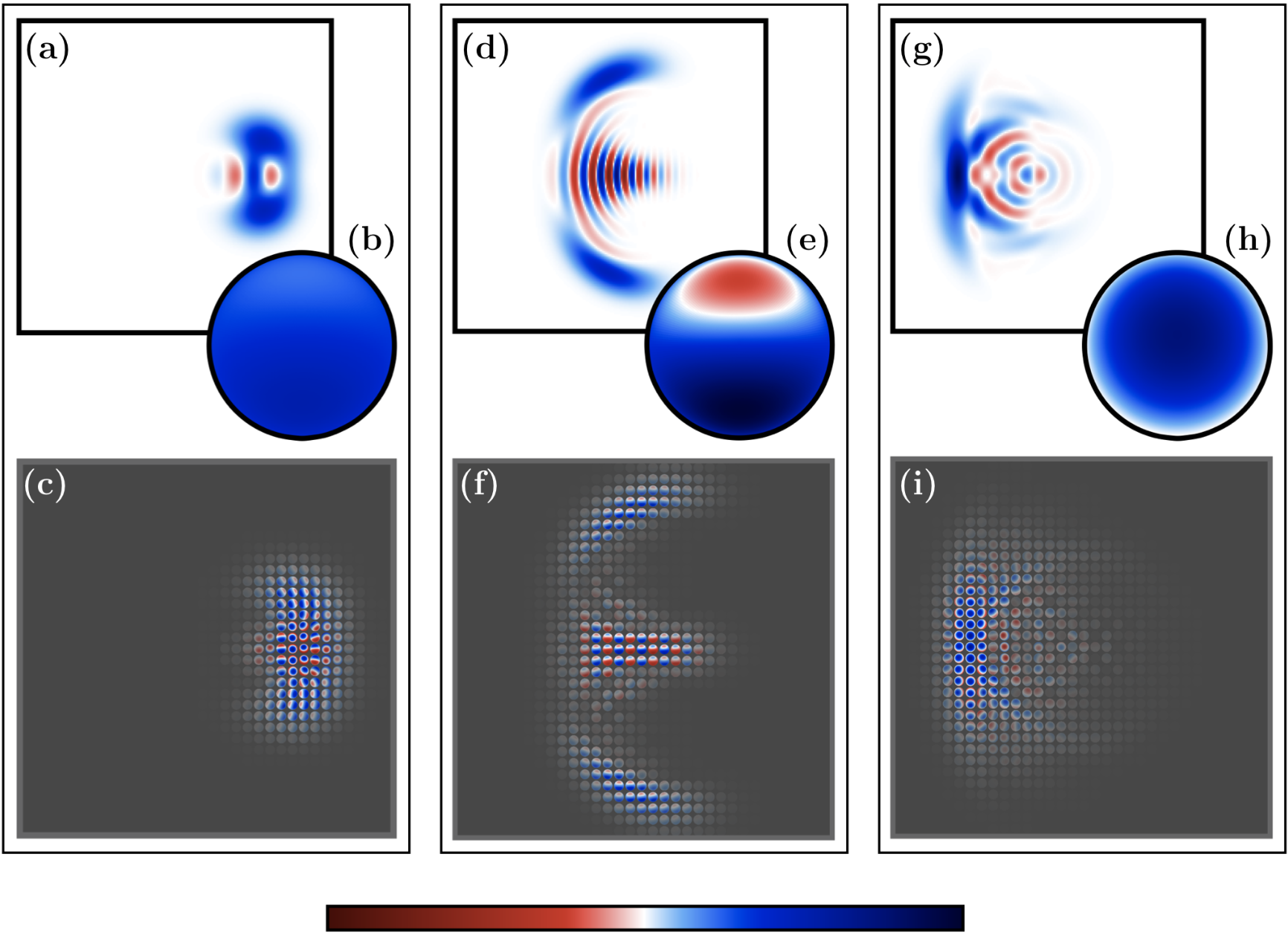}
\caption{\label{FIG:JCM} Here we show the Wigner functions for three points in the evolution of the Jaynes-Cummings model with initial state $\ket{\beta}_f\Excited_a$, where $\beta = 3$.
The reduced Wigner functions are given on the top, where the reduced CV Wigner functions are in \Sub{a}, \Sub{d}, and \Sub{g}.
The reduced DV Wigner functions are in \Sub{b}, \Sub{e}, and \Sub{h}.
The hybrid Wigner functions for the coupled system are in \Sub{c}, \Sub{f}, and \Sub{i}.
The values for the colours correspond to the same values in~\Fig{ExampleSeparableCats} and~\Fig{ExampleEntangledCats}.}
\end{figure}

We now consider the JCM evolution with a different initial state. 
The vacuum state is replaced by a coherent state, giving the initial state $\ket{\beta}_f\Excited_a$, where again $\beta = 3$.
This choice of intial state produces very different effects in the Jaynes-Cumming model, such as the collapse and revival of the Rabi oscillations, where the revival of the Rabi oscillations happen at time $t_r$.
Three noteworthy snapshots, points within the evolution, of the Jaynes-Cummings model are shown in~\Fig{FIG:JCM}.
For each of the snapshots the reduced Wigner functions are~\FigsSub{FIG:JCM}{a},~\Sub{d}, and~\Sub{g} for the CV system, and~\FigsSub{FIG:JCM}{b},~\Sub{e}, and~\Sub{h} for the DV qubit.
In~\FigsSub{FIG:JCM}{c},~\Sub{f}, and~\Sub{i} are the full Wigner function for each of these snapshots.

The first snapshot is early on in the evolution, $t\approx t_r/9$, where there is a high degree of coupling between the two systems.
The reduced Wigner functions in~\FigsSub{FIG:JCM}{a} and~\Sub{b}, indicate that something approaching a Schr\"odinger cat state forming in the CV system; where the DV qubit is in a highly mixed state.
All that can be deduced from the reduced Wigner functions then is that there are correlations between the qubit and the field mode; the nature of the quantum correlations remains hidden.

Evaluating the full Wigner function in~\FigSub{FIG:JCM}{c}, a better appreciation of the quantum correlations at this point in the evolution can be obtained.
The DV spin direction at the top of the CV Wigner functions are orthogonal to those in the bottom of the CV Wigner function.
Where at the top, the spins point in the direction of the negative eigenstate of $\Sx$; at the bottom they point in the positive eigenstate of $\Sx$.
The correlations found in the middle in~\FigSub{FIG:JCM}{c} match the quantum correlation signature for a coherent state qubit, as they are of a form similar to the traceless states in~\Fig{FIG:SPINS}.

The second snapshot of the Jaynes-Cummings model, $t\approx t_r/2$, is where the field mode and the qubit disentangle, transferring the quantum correlations to form a CV Schr\"odinger cat state.
Presence of this Schr\"odinger cat state is immediately visible in the reduced CV Wigner function in~\FigSub{FIG:JCM}{d}.
The reduced DV Wigner function in~\FigSub{FIG:JCM}{e} has now increased in both negative and positive amplitudes, rotating to the eigenstate of $\Sy$ with eigenvalue $-1$.
The return of coherence of the DV qubit is a good indication that the correlations between the two systems have decreased.

Both of the reduced Wigner functions in~\FigsSub{FIG:JCM}{d} and~\Sub{e} suggest that this state is similar to the example state in~\Fig{ExampleSeparableCats}, which is approximately separable.
Observation of \FigSub{FIG:JCM}{f} confirms this suggestion, but more detail can still be found.
Although very few correlations appear between the two subsystems, some residual quantum correlation has remained between the two.
These correlations are found in the slight twisting of the qubits around the two cats and within the quantum correlations in between.

The final snapshot occurs at the revival of the Rabi oscillations, $t\approx t_r$, where the qubit state is closest to the initial state within the revival.
In~\FigSub{FIG:JCM}{h} the average spin is pointing in the direction of an excited state $\Excited_a$, however, there is a loss of coherence associated with the decrease in amplitudes and no negative values.
The full Wigner function reveals why the coherences in the reduced DV Wigner function have formed.
At most points in the full Wigner function, the DV Wigner function is in the excited state, however at many points there are rotations in the qubit Wigner functions, indicating some residual quantum correlations.
The strongest coherent states are found on the left-hand side, where it appears the state is returning to the initial state of a coherent state coupled to $\Excited_a$. 

The quantum correlations that accompany the two choices of CV qubits have a somewhat different nature however their signatures are distinguishable when considering the full Wigner function.
The correlations for the Fock state qubits show a dependence on each other, arising due to the non-separability of the state. 
This closely resembles the pattern found in spin-orbit coupled states~\cite{Davies2018}, and is comparable to spin texture images.
The fundamental signatures come from the behaviour of the coherences and correlations within and between the systems.
The form of the Wigner function of a two-mode squeezed state, although lacking negative values due to it being Gaussian, resembles the signature identified for the Bell-Fock states; the spatial dependence of one system affecting the state in the other system.

\section{Conclusions}
By plotting the information generated by calculating the Wigner function for a CV-DV hybrid system, we have shown that the usual techniques for visualizing these systems misses the full nature of the quantum correlations that arise.
In turn, our visualization techniques allow us to characterize signatures of quantum correlations that can be found in certain systems; a result that promises potential usefulness in analyzing the correlations in maximally entangled states and entanglement as a result of squeezing.

Moreover, we have shown that by plotting the full Wigner function, we can visually determine the level of quantum correlations that are not always clear in coupled systems, giving clear advantage over reduced Wigner function methods that do not always detect the purity of Bell-cat like entanglement.

By demonstrating these methods within the Jaynes-Cummings model, we show how excitations are shared and swapped, demonstrating a visual representation of the transfer of quantum information between systems.
Extending these methods to different systems, will allow for a more intuitive picture of how quantum information moved around coupled systems, providing further insight into the inner process of quantum processes and algorithms.

There have been previous experimental examples which have used phase space to investigate the types of state considered in this paper. 
One notable example is \Ref{PhysRevLett.109.240501}, from a sequence of measurements of the expectation values of the qubit in different bases, they have been able to recreate the CV Wigner function. Using a similar procedure with our generalized displaced parity operator, it should be possible to extend this to produce experimental results equivalent to those in this paper. This technique could be considered to be a form of quantum state spectroscopy.

\ack
RPR is funded by the EPSRC [grant number EP/N509516/1]. 
TT notes that this work was supported in part by JSPS KAKENHI (C) Grant Number JP17K05569. 
The authors would like to thank Kae Nemoto, William Munro and T~D~Clark for interesting and informative discussions.

\section*{References}
\bibliographystyle{unsrt}
\bibliography{refs}  

\end{document}